\documentclass{article}

\usepackage{mylogic,amsmath,amssymb}

\newcommand{\R}{{\mathbb R}}
\newcommand{\Q}{{\mathbb Q}}

\newcommand{\Z}{{\mathbb Z}}

\title{A decision procedure for linear ``big O''
  equations%\footnote{This is a DRAFT.}
}

\author{Jeremy Avigad and Kevin Donnelly}

\begin{document} 

\maketitle

\begin{abstract}
  Let $F$ be the set of functions from an infinite set, $S$, to an
  ordered ring, $R$. For $f$, $g$, and $h$ in $F$, the assertion $f =
  g + O(h)$ means that for some constant $C$, $|f(x) - g(x)| \leq C
  |h(x)|$ for every $x$ in $S$. Let $L$ be the first-order language
  with variables ranging over such functions, symbols for $0, +, -,
  \min, \max$, and absolute value, and a ternary relation $f = g +
  O(h)$. We show that the set of quantifier-free formulas in this
  language that are valid in the intended class of interpretations is
  decidable, and does not depend on the underlying set, $S$, or the
  ordered ring, $R$. If $R$ is a subfield of the real numbers, we can
  add a constant $1$ function, as well as multiplication by constants
  from any computable subfield. We obtain further decidability results
  for certain situations in which one adds symbols denoting the
  elements of a fixed sequence of functions of strictly increasing
  rates of growth.
\end{abstract}

%=========================================================================

\section{Introduction}
\label{introduction:section}

Let $F$ be the set of functions from any infinite set $S$ to
any ordered ring $R$, and let $f, g, h, \ldots$ range over elements of
$F$. The assertion $f = O(g)$, read ``$f$ is big O of $g$,'' means
that there is a constant $C$ such that for every $x$, $|f(x)| \leq C
|g(x)|$. More generally, the assertion $f = g + O(h)$ means that $f -
g = O(h)$; in other words, there is a constant $C$ such that for every
$x$,
\[
|f(x) - g(x)| \leq C |h(x)|.
\]
Read this as saying that $f$ and $g$ have the same rate of growth up
to that of $h$. The notion is used widely in mathematics and computer
science as a means of characterizing functions and their behaviors. 

Determining the validity of entailments between big O equations
involving even only linear expressions can be tricky. For example, the
entailments 
\[
\left.
\begin{aligned}
f + g & = h + O(k) \\
g + l & = h + O(k)
\end{aligned}
\right\} \Rightarrow f = l + O(k)
\]
and 
\[
\left.
\begin{aligned}
f + g & = h + O(k) \\
g & = O(l) \\
k & = O(l)
\end{aligned}
\right\} \Rightarrow f = h + O(l)
\]
follow from the definitions above. Proofs in analysis often involve
long sequences of such calculations based on facts like these. This is
the case in analytic number theory; infrastructure for big O
calculations was needed to support the formal verification of an
elementary proof of the prime number theorem
\cite{avigad:donnelly:04,avigad:et:al:unp} using the proof assistant
Isabelle~\cite{nipkow:et:al:02}. See also Graham et
al.~\cite{graham:et:al:94} for a helpful overview of big O notation
and its properties.

Let $L$ be the first-order language with variables $f, g, h, \ldots$,
symbols for $0, +, -, \min, \max$, and absolute value, and a ternary
relation $f = g + O(h)$. We show that the set of quantifier-free
formulas in this language that are valid in the intended class of
interpretations is decidable, and does not depend on the underlying
set, $S$, or the ordered ring, $R$. When $S$ itself has an ordering,
$f = g + O(h)$ is sometimes read as the assertion that $f$ and $g$
\emph{eventually} have the same rate of growth up to $O(h)$, that is,
that for some $C$ and $d$, $|f(x) - g(x)| \leq C |h(x)|$ for all $x
\geq d$. We show that this reading of big O equations does not change
the set of valid formulas. If $R$ is a subfield of the real numbers,
we can add a constant $1$ function, as well as multiplication by
constants from any computable subfield.

In fact, we even have decidability in certain situations where we add
a sequence of function symbols $\la g_\alpha \ra$, indexed by elements
$\alpha$ of a computable ordering $I$, denoting a fixed sequence of
functions with strictly increasing rates of growth. For example,
suppose we are interested in functions from positive integers to the
real numbers. Consider the set of terms built up from variables and
symbols for arbitrary products of the fixed functions
\[
1, \ldots, (\log x)^q, \ldots, x^q, \ldots
e^{qx^r}, \ldots,
\]
where $q$ and $r$ range over rational numbers, using rational linear
combinations, min, max, and absolute value (but neither multiplication
nor composition). Consider the set of Boolean combinations of big O
expressions involving these terms that are valid when $f = g + O(h)$
is interpreted as the assertion that $f$ and $g$ eventually have the
same rate of growth up to $O(h)$. We show that this set is decidable.

In practice, big O reasoning is often used when the terms involve sums
of functions that take only nonnegative values. Handling this case is
somewhat easier than the more general one. Our strategy is therefore
to deal with that case first, and then reduce the general case to the
more restricted one. In both cases, big O relations are transitive: if
$r = s + O(t)$ and $t = O(u)$, then $r = s + O(u)$. In the more
restricted case, two equations $r_1 = s_1 + O(t_1)$ and $r_2 = s_2 +
O(t_2)$ entail their sum, $r_1 + r_2 = s_1 + s_2 + O(t_1 + t_2)$, and
$f_1 + \ldots + f_k = O(t)$ entails $f_i = O(t)$ for each $i$. Also, a
variable need only appear once inside the $O$; for example, $O(f + f)$
is the same as $O(f)$. Below, we will show, roughly, that all valid
entailments are obtained in this way. Thus, our decision procedure
works by using these principles to derive consequences from a set of
hypotheses until a saturation point is reached; an equation $r = s +
O(t)$ then follows from the hypotheses if and only if $r = s$ is a
linear combination of the equations that have been determined to hold
up to $O(t)$.

% The set of functions from any set to a ring again forms ring, with the
% constant function $0$ and $1$ and addition and multiplication defined
% pointwise. More generally, $f = g + O(h)$ denotes the fact that the
% absolute value of $f - g$ is big O of $h$, i.e.~that $f$ and $g$ look
% roughly the same, ``up to a multiple of $h$.''

% Calculations with $O$ equations can require some thought. For
% example, from the equations
% together with background knowledge \ldots one can conclude. The
% question we address here is this: is there an algorithm to decide, in
% general, when such an implication is valid?

% We show that the answer is ``yes'' when the equations are of the form
% $r = s + O(t)$, where $r$, $s$, and $t$ are finite sums of function
% variables. Indeed, we show that the question does not depend on the
% ordered ring in question, so the same general entailments hold, say,
% for functions from $\N$ to $\N$ and for functions from an interval to
% the nonnegative reals. We also provide a set of axioms that suffices
% to establish any such valid entailment.

It should not be difficult to incorporate variants of our algorithms
to support formal verification with mechanized proof assistants such
as ACL2~\cite{kaufmann:et:al:00}, Coq~\cite{bertot:casteran:04},
HOL~\cite{gordon:melham:93}, Isabelle~\cite{nipkow:et:al:02}, or
PVS~\cite{owre:et:al:92}. These algorithms cover a large number of
straightforward big O inferences that were used to verify the prime
number theorem. (They do not cover, however, inferences that involve
multiplicative properties of big O reasoning; see the discussion in
Section~\ref{questions:section}.) We therefore view the
questions addressed here as an example of the kinds of interesting
theoretical issues that can emerge from such efforts, and the
resulting algorithm as an example of the kinds of domain-specific
support that can be useful.

We are grateful to two anonymous referees for many corrections and
improvements, and to one of them for finding a problem with our
initial formulation of the results in Section~\ref{sequence:section}.

%=========================================================================

\section{An axiomatization of positive big O equations}
\label{axioms:section}

The simplest version of our decision procedure acts on expressions in
the following language, $L$, for first-order logic with equality:
terms are built up from variables $f_1, f_2, \ldots$ and a constant
symbol, $0$, using a binary function symbol, $+$, and there is one
ternary relation in the language, written $r = s + O(t)$.

In the intended class of interpretations, the variables range over
functions $f_1, f_2, \ldots$ from a set $S$ to an ordered semiring,
that is, the nonnegative part of an ordered ring $R$. We assume that
the ring is nontrivial, so zero is not equal to one. The symbol $+$
denotes pointwise addition, $0$ denotes the constant zero function,
and $f = g + O(h)$ denotes the assertion that there is a $C$ in the
ring such that $| f(x) - g(x) | \leq C |h(x) |$ for all $x$ in
$S$.\footnote{It is common to define $f = O(g)$ to mean $f \in O(g)$,
  where $O(g)$ is defined to be the set of functions $f$ satisfying
  $\fa x (|f(x)| \leq C | g(x) |)$ for some $C$. The expression $f = g
  + O(h)$ is then defined to mean $f - g = O(h)$. These definitions
  are clearly equivalent to the ones we have presented. While it can
  be convenient to use the set formulation when formalizing such
  notions in higher-order logic, the formulations we use have the
  virtue of being first-order.

  Big O notation also makes sense for functions from a set to an
  ordered group; see the discussion at the end of
  Section~\ref{procedure:section}.}

Below we provide a list of axioms, whose universal closures are true
for set $F$ of functions in the intended interpretation. Here, we are
only concerned with the quantifier-free consequences of these
axioms. By Herbrand's theorem, a quantifier-free formula is provable
from universal axioms using first-order logic with equality if and
only if there is a propositional proof of that formula from finitely
many instances of the axioms, together with instances of equality
axioms.  So, instead of a first-order proof system, we can just as
well consider the quantifier-free proof system whose nonlogical axioms
consist of all the instances of the formulas below.

We will write $r =
O(s)$ instead of $r = 0 + O(s)$. In the second-to-last axiom, the
notation $k f$ abbreviates a sum $f + f + \ldots + f$ of $k$ many
$f$'s. The axioms are as follows.
\begin{enumerate}
\item $f = g \liff f = g + O(0)$
\item axioms asserting that $+$ is associative and commutative, with
  identity $0$
\item axioms asserting that for fixed $h$, the relation $f = g + O(h)$
  is reflexive, symmetric, and transitive
\item monotonicity: $f = O(f + g)$
\item transitivity: $f = g + O(h) \land h = O(k) \limplies f = g +
  O(k)$
\item linearity:
\begin{enumerate}
\item $f_1 = g_1 + O(h) \land f_2 = g_2 + O(h) \limplies f_1 + f_2 =
  g_1 + g_2 + O(h)$
\item $f_1 + f_2 = g_1 + g_2 + O(h) \land f_1 =
  g_1 + O(h) \limplies f_2 = g_2 + O(h)$
\item for each positive integer $k$, the axiom $k f = k g + O(h)
  \limplies f = g + O(h)$
\end{enumerate}
\end{enumerate}
The first axiom implies that the equality symbol can be eliminated in
favor of equality ``up to $O(0)$.'' The transitivity axiom asserts
that if $r = O(s)$, then any equation that holds up to $O(r)$ also
holds up to $O(s)$. Thus a relation of the form $r = O(s)$ induces an
inclusion on the set of equations that hold up to $O(r)$ and $O(s)$,
respectively.

Let us consider some consequences of the axioms. First, monotonicity
and transitivity imply
\[
f + g = O(h) \limplies f = O(h).
\]
Intuitively, this is clear, since we have $f \leq f + g$.  Also,
monotonicity, transitivity, and the first linearity axiom yield a
slightly stronger form of linearity:
\[
f_1 = g_1 + O(h_1) \land f_2 = g_2 + O(h_2)
  \limplies f_1 + f_2 = g_1 + g_2 + O(h_1 + h_2).
\]
The third linearity axiom then implies that for any positive integers
$k_1, \ldots, k_m$,
\[
k_1 f_1 + \ldots k_m f_m = O(f_1 + \ldots + f_m).
\]
Of course, we also have $f_1 + \ldots + f_m = O(k_1 f_1 + \ldots k_m
f_m)$. It is convenient to express these last two facts by writing
$O(f_1 + \ldots + f_m) = O(k_1 f_1 + \ldots k_m f_m)$. This means that
a rate of growth $O(t)$ only depends on the variables that appear in
$t$, and not the number of times that they occur.

If $f = O(t)$, linearity implies $s + f = s + O(t)$. Thus if $s'$
denotes the result of deleting occurrences of $f$ in $s$, then $f =
O(t)$ implies $s = s' + O(t)$. This means that in an equation $r = s +
O(t)$, all that is relevant are the variables appearing in $t$, and
the parts of $r$ and $s$ that do not involve variables in $t$. In
other words, if $t'$ denotes the sum of the distinct variables
occurring in $t$, and $r'$ and $s'$ denote the result of deleting
these variables from $r$ and $s$, respectively, then $r = s + O(t)$ is
equivalent to $r' = s' + O(t')$. For example,
\[
3 f_1 + 2 f_2 = 5 f_3 + O(f_2 + 3 f_4)
\]
is equivalent to 
\[
3 f_1 = 5 f_3 + O(f_2 + f_4).
\]
Moreover, $f = O(t)$ implies $O(t) = O(t + f)$. So deriving equations
of the form $f = O(t)$ can both enlarge the set of equations that are
known to hold up to $O(t)$ by adding any equations that are known to
hold up to $O(t + f)$, and simplify equations are that already known
to hold up to $O(t)$ by making $f$ irrelevant. Note, finally, that for
any term $s$, $f + s = O(t)$ implies $f = O(t)$. This means that we
can derive equations of the form $f = O(t)$ by finding a linear
combinations of equations that are known to hold up to $O(t)$ that
result in an equation of the form $f + s = O(t)$.

It will be convenient below to work with big O equations of the form
\begin{equation}
\label{linear:eq}
a_1 f_1 + \ldots + a_m f_m = O(t)
\end{equation}
where $a_1, \ldots, a_m$ are arbitrary \emph{rational}
coefficients. Negative values can easily be interpreted away by moving
the terms to the other side of the equation; for example, $3 f_1 - 2
f_2 = O(f_3)$ can be viewed as an abbreviation for $3 f_1 = 2 f_2 +
O(f_3)$. Similarly, equations involving fractional coefficients can be
understood in terms of the result of multiplying through by the least
common multiple. Of course, for implementation purposes, one should
take these equations at face value, rather than treating them as
metamathematical abbreviations for much longer expressions.

Now suppose we are given a system of equations
\begin{equation}
\label{linear:b:eq}
a_{i,1} f_1 + \ldots + a_{i,m} f_m = O(t)
\end{equation}
for fixed $t$ and $i = 1,\ldots,n$. The linearity axioms imply that
any linear combination of the expressions on the left-hand side also
has rate of growth $O(t)$. Thus we can use conventional methods of
linear algebra to derive new equations of the form (\ref{linear:eq}).

%=========================================================================

\section{A combinatorial lemma}
\label{combinatorial:section}

Let us consider where we stand. With helpful notational abbreviations,
we have focused our attention on formulas of the form
(\ref{linear:eq}), where the coefficients are rational
numbers. Without loss of generality, we can assume $t$ is a sum of
distinct variables, and that these variables are disjoint from $f_1,
\ldots, f_m$. Suppose we start with a set of hypotheses and derive a
set of equations of the form (\ref{linear:b:eq}), for a fixed $t$,
with $i=1,\ldots,n$. We can both enlarge and simplify this set of
consequences by deriving new formulas $f_v = O(t)$ for $v =
1,\ldots,m$. We can do that, in turn, by finding linear combinations
of the equations (\ref{linear:b:eq}) that yield formulas of the form
(\ref{linear:eq}) in which each $a_i$ is nonnegative and $a_v$ is
strictly positive for some $v$.

In this section, we show that it is algorithmically decidable whether
such a linear combination of the equations exists. We will also
provide a dual characterization of this condition that will ultimately
enable us to show that our decision procedure for quantifier-free big
O expressions is complete. The decision procedure itself will be
presented in the next section.

Suppose we are given a system of $n$ equations of the form
(\ref{linear:b:eq}), where $i$ runs from $1$ to $n$. A rational linear
combination of the expressions on the left-hand-side is an expression
of the form
\begin{equation}
\label{b:eqn}
\sum_{i = 1\ldots n} b_i a_{i,1} f_1 + \ldots + \sum_{i = 1 \ldots n}
b_i a_{i,m} f_m
\end{equation}
for some sequence of rational numbers $b_1, \ldots, b_n$. We would
like to know whether there is a choice of $b_1, \ldots, b_n$ that
makes all the coefficients nonnegative, and at least one coefficient
strictly positive.

Let $A$ be the $n \times m$ matrix of rational numbers $\la a_{i,j}
\ra_{i = 1\ldots n, j = 1 \ldots m}$. If we use $f$ to denote the
vector of variables $\la f_1, \ldots, f_n \ra$, and we let $f^t$
denote its transpose, then the equations (\ref{linear:b:eq})
are just the rows of $A f^t$. If $b$ is the vector $\la b_1, \ldots,
b_n \ra$, then $b A f^t$ is expression (\ref{b:eqn}), and $b A$ is the
vector of the $m$ coefficients.

\begin{lemma}
\label{first:lemma}
  Let $A$ be an $n \times m$ matrix of rational numbers, and let $v$
  be any index, $1 \leq v \leq m$. Then the question as to whether
  there is any vector $b = \la b_1, \ldots, b_n \ra$ such that $b A$
  is nonnegative and the $v$th element is strictly positive is
  decidable.
\end{lemma}

\proof This is a system of $m$ inequalities in $n$ unknowns, and so
the problem amounts to determining whether a linear program is
feasible. This is easily solved using standard linear programming
techniques \cite{apt:03,papadimitriou:steiglitz:98}.  \proofend

In Section~\ref{procedure:section}, we will use the following dual
characterization of the problem.

\begin{lemma}
  Let $A$ be an $n \times m$ matrix of rational numbers, and let $v$
  be any index, $1 \leq v \leq m$. Then the following
  two conditions are equivalent:
\begin{enumerate}
\item There is a vector $b = \la b_1, \ldots, b_n \ra$ such that $b A$
  is nonnegative, and the $v$th component of $b A$ is strictly
  positive.
\item There is no nonnegative vector $f = \la f_1, \ldots, f_m \ra$ of
  rational numbers satisfying $A f^t = 0$ and $f_v > 0$.
\end{enumerate}
\end{lemma}

\proof To see that \emph{1} implies \emph{2}, suppose \emph{2} is false. Then there is a
nonnegative vector $f = \la f_1, \ldots, f_m \ra$ of rational numbers
with $A f^t = 0$ and $f_v > 0$. Then $b A f^t = 0$ for every $b$, that
is, the expression $\sum_{i = 1\ldots n} b_i a_{i,1} f_1 + \ldots +
\sum_{i = 1 \ldots n} b_i a_{i,m} f_m$ is equal to $0$. If, on the
other hand, \emph{1} holds, there is a $b$ such that each term of this
expression is nonnegative and the $v$th summand is strictly positive,
making the expression strictly positive. Thus if \emph{2} is false, \emph{1} is
false as well.

The fact that \emph{2} implies \emph{1}, and, in fact, the full equivalence, is a
direct consequence of the duality theorem for linear
programming. Consider the following two problems:
\begin{enumerate}
\item Find a vector $b$ maximizing the constant function 0, subject to
  the constraints $b A \geq \la 0, 0, \ldots, 0, 1, 0, \ldots, 0 \ra$,
  where the $1$ occurs in the $v$th position.
\item Find a vector $f$ minimizing $-f_v$, subject to the constraints
  $f \geq 0$ and $A f^t = 0$.
\end{enumerate}
By the duality theorem (\cite[Theorem 3.1]{papadimitriou:steiglitz:98}
or \cite[Theorem 8.3.1]{hall:86}), the first problem has a solution if
and only if the second one does.

Now suppose there is a $b$ such that each component of $b A$ is
nonnegative, and the $v$th component is strictly positive. Scaling $b$
by the reciprocal of the $v$th component, we get a vector $b'$ such
that $b' A$ is nonnegative and the $v$th component is greater than or
equal to $1$. Thus the first problem has a solution if and only if
condition \emph{1} of the lemma holds.

On the other hand, $A f^t = 0$ has at least one solution, namely, when
$f$ is the constant $0$ vector. Suppose $f$ is a nonnegative vector
such that $A f^t = 0$ and $f_v$ is strictly positive. Then any
multiple of $f$ also has this property, and the multiples of $-f_v$
are unbounded. Thus the second problem has a solution if and only if
for every $f$ satisfying $A f^t = 0$ and $f \geq 0$, we have $f_v =
0$; that is, if and only if condition \emph{2} of the lemma holds. So
the two conditions are equivalent, as claimed. \proofend

The following fact will also be useful in proving completeness.

\begin{lemma}
\label{third:lemma}
Let $A$ be an $n \times m$ matrix of rational numbers, and suppose for
every $v$ from $1$ to $m$ there is a nonnegative vector $f$ such that
$A f^t = 0$ and the $v$th component of $f$ is strictly positive. Then
there is a vector $f$ such that $A f^t = 0$, and every component of $f$
is strictly positive.
\end{lemma}

\proof For each $v$, choose a vector $f_v$ satisfying the
hypothesis. Then the sum $f = \sum_{v=1}^m f_v$ of these vectors
satisfies $A f^t = \sum_{v = 1}^m A f_v^t = 0$, and every component of
$f$ is strictly positive. \proofend

%=========================================================================

\section{A decision procedure}
\label{procedure:section}

Let $L$ be the language described in Section~\ref{axioms:section}. Let
$S$ be any set, let $R$ be any ordered ring, and let $F$ be the set of
functions from $S$ to the nonnegative part of $R$. Say that a
quantifier-free formula in $L$ is \emph{valid in $F$} if its universal
closure holds in $F$, that is, if the formula is true for all
instances of the variables under the intended interpretation.

Before considering arbitrary quantifier-free formulas, we first
consider \emph{Horn clauses}. These are formulas of the form
\[
\ph_1 \land \ldots \land \ph_k \limplies \psi
\]
where each $\ph_i$ and $\psi$ is an atomic formula. We will prove:
\begin{theorem}
\label{main:theorem}
  Let $L$ and $F$ be as above. The set of Horn clauses that are valid
  in $F$ is decidable, and do not depend on the choice of $S$ or $R$.
\end{theorem}
In particular, the valid Horn clauses are exactly the ones that hold
of the set of functions mapping a single element to the natural
numbers.

Now consider any quantifier-free formula in $L$. Classically, this
formula is equivalent to one in conjunctive normal form, that is, a
conjunction of disjunctions of literals (i.e.~atomic formulas and
their negations). A conjunction of formulas is valid in $F$ if and
only if each conjunct is valid in $F$, so to provide a decision
procedure for arbitrary quantifier-free formulas, it suffices to
provide a decision procedure for disjunctions of literals. But any
such disjunction is equivalent to a formula of the form
\begin{equation}
\label{disjunction:eq}
\ph_1 \land \ldots \land \ph_k \limplies \psi_1 \lor \ldots \lor
\psi_l,
\end{equation}
where each $\ph_i$ and $\psi_j$ is an atomic formula, this is, a big O
equation. If any of the implications 
\begin{equation}
\label{imp:eq}
\ph_1 \land \ldots \land \ph_k \limplies \psi_j
\end{equation}
is valid in some $F$ (and so, by Theorem~\ref{main:theorem}, in all
$F$'s), then clearly (\ref{disjunction:eq}) is valid in all $F$'s. On
the other hand, if there is a counterexample to each equation
(\ref{imp:eq}), then by Theorem~\ref{main:theorem} there is a
counterexample consisting of a function from a singleton to the
natural numbers. We can combine these $l$ counterexamples into a
single counterexample consisting of functions from $\{ 1, \ldots, l
\}$ to $\N$, where each variable $f$ is interpreted as the function that
takes the value of the $j$th counterexample on input $j$. This
provides a counterexample to (\ref{disjunction:eq}). Since there is no
structure on the set $S$, all that matters is its cardinality; so we
have that the formula (\ref{disjunction:eq}) is valid for all $F$'s
for which $S$ is sufficiently large if and only if each Horn clause
(\ref{imp:eq}) is valid in every $F$. So Theorem~\ref{main:theorem}
has the following consequence.
\begin{theorem}
\label{main2:theorem}
Let $F$ be the set of functions from any infinite set $S$ to the
nonnegative part of any ordered ring $R$. Then the set of
quantifier-free formulas that are valid in $F$ is decidable, and does
not depend on $S$ or $R$.
\end{theorem}
If $S$ is an ordered set with no greatest element, one sometimes finds
alternative readings of $r = s + O(t)$ to the effect that the rate of
growth is bounded \emph{eventually}, that is, for all suitably large
$x$. (If $S$ has a greatest element, the notion degenerates, depending
on whether one uses $>$ or $\geq$ to express ``suitably large.'') Once
again, a decision procedure for arbitrary quantifier-free formulas
reduces to a decision procedure for Horn clauses. It is not hard to
verify that if a Horn clause is valid under the original reading, it
is valid under the ``eventually'' reading. Conversely, it is not hard
to turn a counterexample to the original reading where the domain $S$
is is a singleton into a counterexample to the ``eventually'' reading
for any ordered $S$ using the corresponding constant functions. So we
have:
\begin{theorem}
The set of quantifier-free formulas of $L$ that are valid for every
set of functions from an ordered set with no greatest element to
the nonnegative part of an ordered ring on the ``eventually'' reading
coincides with the set of formulas named in
Theorem~\ref{main2:theorem}.
\end{theorem}

\proofarg{of Theorem~\ref{main:theorem}} We will describe an algorithm
for determining whether a Horn clause is valid, and show that the
algorithm behaves as advertised. Suppose we are given a Horn clause
with variables among $f_1, \ldots, f_m$. Without loss of generality we
can assume that the hypotheses are all of the form $q = O(r)$, where
$q$ is a rational linear combination of $f_1, \ldots, f_m$, and $r$ is
a sum of distinct variables from among $f_1, \ldots, f_m$. We can also
assume that the conclusion, $s = O(t)$, is of this same form. Our task
is to decide whether the conclusion is entailed by the hypotheses.

For any subset $A$ of $\{ f_1, \ldots, f_m \}$, it will be convenient
to write $t_A$ for the sum $\sum_{f_i \in A} f_i$ of the variables in
$A$. Also, if $q$ is a rational linear combination of $f_1, \ldots,
f_m$, it will be convenient to write $q[A]$ for the result of setting
the coefficient of $f_i$ to zero for each $f_i$ in $A$. We saw in the
previous section that for any $s$ and $t$, if $A$ is the set of
variables occurring in $t$, then $s = O(t)$ is equivalent to $s[A] =
O(t_A)$. Also, if the indices of the variables of $r$ are all in $A$,
then $q = O(r)$ entails $q = O(t_A)$, which is equivalent to $q[A] =
O(t_A)$.

The algorithm is as follows:
\begin{quote}

  Set $A$ equal to the set of variables occurring in $t$.

Repeat:

\begin{quote}

  Let $Q$ be the set of terms $q[S]$ where $q = O(r)$ is a hypothesis
  and the variables of $r$ are all in $A$.

For each $f_v \in \{ f_1, \ldots, f_m \} - A$:

\begin{quote}
If there is a rational linear combination of elements of $Q$ with
nonnegative coefficients and positive $v$th coefficient, add $f_v$ to
$A$. 
\end{quote}

\end{quote}

until no new indices are added to $A$.

  Let $Q$ be the set of terms $q[S]$ where $q = O(r)$ is a hypothesis
  and the variables of $r$ are all in $A$.

If $s[A]$ is a linear combination of elements of $Q$, return
``true,'' else return ``false.''

\end{quote}

We start by setting $A$ to be the set of variables occurring in $t$,
so $O(t) = O(t_A)$. At each pass through the outer loop, we try to
augment $A$ while maintaining $O(t) = O(t_A)$. Suppose we have a
hypothesis $q = O(r)$, where the variables of $r$ are all in $A$. Then
$r = O(t_A)$. By transitivity, we have $q = O(t_A)$, which is
equivalent to $q[A] = O(t_A)$. Thus we let $Q$ be the set of terms
$q[A]$ corresponding to such $r$. Then any linear combination of
elements of $Q$ also has order of growth $O(t_A)$. If some such linear
combination has nonnegative coefficients, and the coefficient of $f_v$
is strictly positive for some $v$, then we know the $f_v =
O(t_A)$. This implies $O(t) = O(t_A) = O(t_A + f_v) = O(t_{A \cup \{
  f_v\} })$, and we add $f_v$ to $A$. The outer loop terminates when
we can no longer derive new expressions of the form $f_v = O(t_A)$.

Once we have left the outer loop, we will have $O(t) = O(t_A)$, and we
once again let $Q$ be the set of terms $q[A]$ such that we have $r =
O(t_A)$. If $s$ is a linear combination of terms in $Q$, then $s =
O(t_A) = O(t)$. Thus we have shown that $s = O(t)$ is a consequence of
the hypothesis in any of the intended interpretations, and we return
``true.'' Otherwise, we return ``false.''

All we have left to do is to show that if the algorithm returns
``false,'' then there is a counterexample in the set of functions $F$
from any set $S$ to the nonnegative part of any ordered ring, $R$. In
fact, we will construct a counterexample where $S = \{ * \}$ is a
singleton and $R$ is the integers. Thus our counterexample amounts to
assigning a nonnegative integer to each variable $f_i$. In that case,
an expression of the form $s = O(t)$ comes out true if and only if $t$
is nonnegative, or $t = 0$ and $s = 0$. Conversely, $s = O(t)$ comes
out false if and only if $t = 0$ and $s$ is strictly positive. Since
every ordered ring contains a copy of the natural numbers and one can
take the corresponding constant functions for any set $S$, this
provides counterexamples for every $S$ and $R$, simultaneously.

We now describe the assignment of nonnegative integers to the
variables $f_i$. Let $A$ be the set of variables at the termination of
the outer loop. For each $f_i$ in $A$, set $f_i = 0$.

We still have to assign values to the variables $f_i$ that are not in
$A$. Let $Q$ be the set of expressions $q[A]$ such that $q = O(r)$ is
one of the hypotheses and the variables of $r$ are in $A$. Since the
outer loop terminates with that value of $A$, by
Lemma~\ref{third:lemma} we know that there is an assignment of
strictly positive rational values $c_i$ to each variable $f_i$ not in
$A$ making each $q[A]$ equal to $0$. Scaling these, we can assume
that each $c_i$ is a strictly positive integer. Also, since $s[A]$ is
not a linear combination of the expressions in $Q$, by linear algebra
there is an assignment of rational values $d_i$ to variables $f_i$ not
in $A$ making each $q[A]$ equal to zero and $s[A]$ nonzero. Scaling
again, we can assume that the values of $d_i$ are integers.

Suppose the value of $s[A]$ under the assignment of the $c_i$'s is $x$
and the value of $s[A]$ under the assignment of the $d_i$'s is
$y$. Since the $c_i$'s are strictly positive and $y$ is nonzero, we
have that for sufficiently large integer $e$, assigning $e c_i + d_i$
to $f_i$ will make $f_i$ strictly positive.  In that case, each $q[A]$
gets the value $0$, and $s[A]$ gets the value $e x + y$. Because $y$
is not zero, we can choose $e$ such that in addition $e x + y$ is not
equal to $0$. So we choose such an $e$ and assign each $f_i$ the value
$e c_i + d_i$.

We need to show that with the assignment of values to the $f_i$'s that
we have just described, each hypothesis $q = O(r)$ comes out true,
while $s = O(t)$ comes out false. First, note that if any variable of
$r$ is not in $A$, then $r$ is strictly positive, and $q = O(r)$ is
true. Thus we only have to worry about hypotheses $q = O(r)$ where
$q[A]$ is one of the expressions in $Q$. In that case, our assignment
of values to $f_i$'s not in $A$ ensures that $q[A]$ has value $0$, and
since we have assigned zero to the other $f_i$'s, we have $q =
q[A]$. Thus each such $q$ has value $0$, and since $0 = O(0)$, the
hypotheses are satisfied.

On the other hand, since the variables of $t$ are all in $A$,
$t$ has a value of $0$ under the assignment. We have also ensured that
the value of $s[A]$, and hence the value of $s$, is strictly
positive. Thus, under the assignment, $s = O(t)$ is false, as
required. \proofend

Note that the inner loop repeats at most $m$ times, where $m$ is the
number of variables occuring in $t$. The bottleneck therefore occurs
in testing the satisfiability of the system of linear inequalities in
the inner loop. This can be done using standard linear programming
techniques \cite{apt:03,papadimitriou:steiglitz:98}. Karmarkar's
algorithm \cite{karmarkar:84}, for example, solves such problems in
time $O(n^{3.5} L \ln L \ln^2 L)$, where $n$ is the number of
variables, and $L$ is the length of the input. This shows that, at
least in principle, our algorithm can be made to run in polynomial
time. In practice, we expect that a simple-minded algorithm like the
Fourier-Motzkin procedure \cite{apt:03} will work quite well, despite
the fact that it can run in double-exponential time in the worst case
\cite{weispfenning:94}. Other methods, such as Dantzig's simplex
method \cite{papadimitriou:steiglitz:98} or Weispfenning's ``test
point'' method \cite{loos:weispfenning:93,weispfenning:88}, are
further options.

We have implemented, in ML, a prototype version of the algorithm just
described, based on the Fourier-Motzkin test. We have confirmed that
it does well on natural examples: on a Pentium M 1.6 GHz processor,
our implementation decides examples with on the order of five or six
variables, like the ones in the introduction, in under 20 ms (which is
about the limit of our timer's precision).

Note that if $R$ is an ordered \emph{group} instead of an ordered
ring, there is still an action of $\Z$ on $R$, taking $k x$ to be a
sum $x + \ldots + x$ of $k$ many $x$'s. Big O notation even makes
sense in this setting, if one interprets the constant $C$ as an
element of $\Z$. The axioms of Section~\ref{axioms:section} are still
valid, and the decision procedure above still works. When $R$ is a
subfield of the real numbers, the two interpretations coincide.

In the other direction, when $R$ is a field, it makes sense to include
multiplication by arbitrary rational constants in the language. Since
the duality principle from linear programming holds for any subfield
$R$ of the real numbers, the procedure also works for such $R$ when we
allow multiplication by constants from any computable subfield, that
is, function symbols $c_a(f) = a f$, for each such $a$.

It is not hard to see that the axioms described in
Section~\ref{axioms:section} are sufficient to prove any entailment
that our procedure sanctions as valid. This yields:
\begin{theorem}
The set of quantifier-free formulas of $L$ valid in the intended class
of interpretations is equal to the set of quantifier-free consequences
of the axioms in Section~\ref{axioms:section}.
\end{theorem}
If we add multiplication by constants, it suffices to add the obvious
identities, like $c_a(f + g) = c_a(f) + c_a(g)$, and so on.

%=========================================================================

\section{Handling negative values}
\label{negative:section}

The absolute value function is defined on any ordered ring by setting
$|x| = x$ if $x \geq 0$, and $|x|=-x$ otherwise. This can be lifted to
functions from a set to an ordered group by defining $|f|$ to be
the function mapping $x$ to $|f(x)|$ for every $x$. 

Let us now extend the language $L$ of Section~\ref{axioms:section} to
a language $L'$ where we add subtraction and absolute value, and now
take the function variables to range over functions from a set $S$ to
an arbitrary ring $R$. The functions $\min$ and $\max$ can then be
defined by the following equations:
\begin{eqnarray*}
\min(f,g) & = & (f + g - |f - g|) / 2 \\
\max(f,g) & = & (f + g + |f - g|) / 2
\end{eqnarray*}
Since $|f|$ is always a nonnegative function and any nonnegative
function can be expressed in this way, the decision procedure in the
previous section can be viewed as working with the fragment of the
language with only addition, and where variables are replaced by
expressions of the form $|f|$. Our goal now is to show that the
procedure extends to the full language.
\begin{theorem}
  Let $F$ be the set of functions from any infinite set $S$ to any
  ordered ring $R$. Then the set of quantifier-free formulas of $L'$
  that are valid in $F$ is decidable, and does not depend on the
  choice of $F$.
\end{theorem}
As before, if $R$ is a subfield of the reals, we can extend the
language with multiplication by constants in any computable subfield.

When functions can take on positive and negative values, the task of
determining what is valid becomes more subtle. The expressions $f_1 =
O(g)$ and $f_2 = O(g)$ still entail $f_1 + f_2 = O(g)$, but it is no
longer necessarily the case that $f = O(g_1)$ and $f = O(g_2)$ entail
$f = O(g_1 + g_2)$, or even that $g_1 = O(g_1 + g_2)$ generally holds:
consider the fact that $g_2$ might be $-g_1$. But if $f$ is any
function, we can subdivide the domain $S$ into a set $S_0$ where the
value of $f$ is nonnegative and a set $S_1$ where the value of $f$ is
nonpositive. In fact, we can do this for all terms appearing in an
expression, creating a partition of $S$ such that on each element of
the partition the signs of the terms do not change. A big O equation
will hold if and if it holds on each segment of the partition, and we
can use this observation to reduce the problem to that which we
solved in Section~\ref{procedure:section}.

In order to spell out the details, we will rely on the following
lemma. We will use variables $\alpha, \beta, \gamma, \ldots$ to range
over nonnegative functions, which can be thought of as expressions of
the form $|a|, |b|, |c|, \ldots$, where $a, b, c, \ldots$ are ordinary
variables of $L'$. From now on we assume we are dealing with functions
from an infinite set $S$ to an ordered ring $R$.

\begin{lemma}
\label{case:split}
Let $\ph(f)$ be any quantifier-free formula in the language of
$L'$. Then $\ph(f)$ is valid if and only if $\ph(\alpha)$ and
$\ph(-\alpha)$ are both valid, where $\alpha$ is a new variable
ranging over nonnegative functions.
\end{lemma}

\proof Clearly if $\ph(f)$ is valid then it holds whenever $f$ is
nonnegative or nonpositive, so $\ph(\alpha)$ and $\ph(-\alpha)$ are
both valid. To verify the converse, as in the previous section, we
only need to consider Horn clauses
\[
\bigwedge q_i = O(r_i) \limplies s = O(t).
\]
So, suppose for some assignment of variables, including the
expression above is false. Then each $q_i = O(r_i)$ is true for this
assignment, but $s = O(t)$ is false. Let $S_0$ be the elements of $S$
where $f$ is nonnegative, and let $S_1$ be $S - S_0$. Then each
hypothesis $q_i = O(r_i)$ remains true when the functions are
restricted to $S_0$ and $S_1$, respectively. Since $s = O(t)$ is
false, it must be false of the restrictions of the functions to either
$S_0$ or $S_1$. As in the previous section, this counterexample on an
$S_i$ can be turned into a counterexample with domain $S$ just by
picking an element $x$ in $S_i$ and setting $f(y) = f(x)$ for $y$ in
$S - S_i$. But now $f$ is either nonnegative or nonpositive, providing
a counterexample to either $\ph(\alpha)$ or $\ph(-\alpha)$.  \proofend

We now describe a procedure for transforming a formula $\ph$ involving
variables $f_1, \ldots, f_m$ into a formula $\ph'$ involving only
variables $\alpha_1, \ldots, \alpha_k$, such that the absolute value
function does not occur in $\ph'$, and such that $\ph$ is valid if and
only if $\ph'$ is. In an expression $s = O(t)$ in $\ph'$, $s$ may be a
rational linear combination of variables, but that can be understood
according the the conventions of Section~\ref{axioms:section}; $t$
will always be a variable, $\alpha$. Thus the decision procedure in
Section~\ref{procedure:section} applies to $\ph'$. 

First, in $\ph$, replace every atomic formula $s = O(t)$ by $s =
O(|t|)$. Clearly, this does not change the interpretation of the
formula.

Now, iteratively, for each expression $|t|$ occurring in $\ph$,
introduce a new variable $h$, add the hypothesis $h = t$,
and replace by $t$ by $h$ in $\ph$. Do this with the innermost
occurrences of $t$ first, so we are left with a formula of the form 
\[
\bigwedge h_i = t_i \limplies \ph,
\]
where the absolute value function does not occur in any $t_i$, and
occurs only in the form $| h_i |$ in $\ph$. 

The result is a formula involving the original variables $f_1, \ldots,
f_m$ of $\ph$, and new variables $h_1, \ldots, h_n$. By
Lemma~\ref{case:split}, this formula is valid if and only if so is the
conjunction obtained by substituting all combinations $\pm \alpha_1,
\ldots, \pm \alpha_{m+n}$ for these variables. Replace $| \pm \alpha_j
|$ by $\alpha_j$, and call the resulting formula $\ph'$. Then $\ph'$
has the requisite form, and we are reduced to
Theorem~\ref{main2:theorem}. \proofend

It is instructive to see how this procedure works on particular
examples. For example, one attempts to verify $f = O(f + g)$ by
considering $f = O(|f + g|)$, and then, in turn, $h = f + g \limplies
f = O(|h|)$. This last formula is valid if every substitution of $\pm
\alpha, \pm \beta$, and $\pm \gamma$ for $f$, $g$, and $h$,
respectively, yields a valid formula. But if we substitute $\alpha$,
$-\beta$, and $\gamma$, we get $\gamma = \alpha -
\beta \limplies \alpha = O(\gamma)$. This is equivalent to $\beta
+ \gamma = O(\gamma)$, which is not generally valid.

Because the procedure involves iterating case splits, the algorithm
runs in exponential time. We do not know whether this upper bound can
be improved. In situations where the signs of subterms are constant
and can be determined, however, such splits can be avoided.

%=========================================================================

\section{Handling constant functions}
\label{constants:section}

In this section, we suppose we are dealing with the set $F$ of
functions from a set $S$ to an ordered field $R$ where there is at
least one function, $G_*$, that does not have constant rate of growth;
i.e.~such that $1 = O(G_*)$ but $G_* \neq O(1)$, where $1$ denotes the
constant function returning one. For example, on functions from $\N$
to $\R$ we can take $G_*(x) = 1 + x$; in general, we can find such a
function as long as there is a cofinal subset of $R$ that has
cardinality at most that of $S$.

We have not included a symbol for the constant function $1$ in the
language of $L$. We can obtain some of the expressions that are valid
in the extended language by using a variable $g_1$ in place of $1$,
and then checking the validity of
\begin{equation}
\label{one:eq}
g_1 \neq O(0) \limplies \ph,
\end{equation}
where $\ph$ is any quantifier-free formula involving $g_1$ and other
variables $f_1, \ldots, f_m$. If this expression is valid, then
clearly $\ph$ is valid when $g_1$ is interpreted as $1$. In this
section we will show, surprisingly, that the converse holds, i.e.~that
\emph{all} valid entailments arise in this way.
\begin{theorem}
  For any quantifier-free formula $\ph$ in the language $L'$, $\ph$ is
  valid when $g_1$ is interpreted as the constant function $1$ if and
  only if the formula
\[
g_1 \neq O(0) \limplies \ph
\]
is valid.
\end{theorem}
As a result, our decidability results hold for the extension to the
the language $L'$ with a symbol to denote the constant one
function. (In structures where $f = O(1)$ holds for every $f$, a
straightforward variation of the decision procedure works.)

\medskip

\proof As before, it suffices to prove the theorem for Horn clauses
and the language $L$, where the variables are assumed to range over
nonnegative functions. Suppose $\ph$ is a Horn clause of the form
$\bigwedge q_i = O(r_i) \limplies s = O(t)$, involving variables
$f_1, \ldots, f_m$ and $g_1$. The formula $g_1 \neq O(0) \limplies
\ph$ is equivalent to
\[
\bigwedge q_i = O(r_i) \limplies g_1 = O(0) \lor s = O(t).
\]
If $\ph$ is not valid, then our algorithm returns ``false'' on both
\[
\bigwedge q_i = O(r_i) \limplies g_1 = O(0).
\]
and
\[
\bigwedge q_i = O(r_i) \limplies s = O(t).
\]
We will show that from this outcome on both runs, we can construct a
counterexample to $\ph$ where $g_1$ is interpreted as $1$.

Since the algorithm returns ``false'' to the first query, we know from
Section~\ref{procedure:section} that there is an assignment of
rational values $c_1, \ldots, c_m, u$ to $f_1, \ldots, f_m, g_1$
making the hypotheses true, but $g_1 \neq 0$. Scaling, we can assume
that $u = 1$. Let $A$ be the set of variables that have been
accumulated by the end of the main loop. Then $A$ is the set of
variables $f$ such that $f = O(0)$ has been determined to be a
consequence of the hypotheses; that is, the set of symbols $f$ such
that we have $f = 0$. We have that $c_i \neq 0$ for each $f_i$ that is
not in $A$. 

Since the algorithm returns ``false'' to the second query, we know
that there is an assignment of rational values to $d_1, \ldots, d_m,
v$ to $f_1, \ldots, f_m, g_1$ making the hypotheses true, and the
conclusion $s = O(t)$ false. In other words, $t$ has a value of $0$,
and $s$ has a nonzero value, under the assignment. Let $B$ be the set
of variables $f$ such that $f = O(t)$ has been determined to be a
consequence of the hypotheses by the end of the second algorithm. Note
that $B$ includes $A$: if $f = O(0)$ is a consequence of the
hypotheses, then so is $f = O(t)$.

Now there are two cases, depending on whether $g_1$ is in the set
$B$ at the end of this second run. If it isn't, then $g_1 = O(t)$ is
not entailed by the hypotheses. In that case, we can proceed as in
Section~\ref{procedure:section}.  The value $v$ assigned to $g_1$ is
strictly positive, so we can scale the assignment so that $v =
1$. Assigning $f_1, \ldots, f_m, g_1$ the constant functions that
return $d_1, \ldots, d_m, v$ provides the desired counterexample. In
this case, we just discard the values $c_1, \ldots, c_m, u$ obtained
from the first run of the algorithm.

Otherwise, the value $v$ assigned to $g_1$ by the second run of the
algorithm is $0$, which is to say, $g_1 = O(t)$ is a consequence of
the hypotheses. In that case, we will construct a
counterexample by assigning functions that are $O(1)$ to variables
$f$ in $A$, that is, the ones that are required to have rate of growth
$O(t)$; and we will assign functions that are $O(G_*)$ to the
rest. Specifically, for each $i$, assign the function $d_i G_* + c_i$
to the variable $f_i$, and assign the function $1 = v G_* + u$ to $g_1$.

Let us show that this works. Consider a hypothesis $q = O(r)$. If $r$
involves any variable $f_i$ not in $B$, then the value of $r$ is
$O(G_*)$, and the hypothesis is automatically satisfied, because all
the functions have growth rate less than or equal to $O(G_*)$.

Otherwise, every $f_i$ occurring in $r$ is in $B$. Suppose for
at least one $f_i$ occurring in $r$, $f_i$ is not in $A$. Then the value
of $r$ is a nonzero constant function. In that case, the value of the
constant terms of the functions assigned to the variables $f_i$ is
irrelevant as to whether the equation is satisfied; all that matters
are the coefficients $d_i$ of $G_*$. But these were chosen by the
second run of the algorithm so that all these hypotheses are
satisfied.

We are left with the case where all the variables occurring in $r$ are
in $A$. In this case, $O(r) = O(0)$ under the assignment. The value of
constant term of $q$ under the final assignment is equal to the value
of $q$ under the assignment of $c_1, \ldots, c_m, u$ to the variables,
and these values were chosen by the first run of the algorithm to
ensure that this is equal to $0$. The value of the coefficient of
$G_*$ in $q$ under the final assignment is equal to the value of $q$
under the assignments of $d_1, \ldots, d_m, v$ to the variables, and
these values were chosen by the second run of the algorithm to ensure
that this is equal to $0$. Thus $q$ is equal to $0$ under the final
assignment.

Finally, we only need to show that $s = O(t)$ comes out false under
the assignment. But we assigned values to the variables of $t$ so as
to ensure that $t$ has value at most $O(1)$, while at the same the
values of $d_1, \ldots, d_m$ guarantee that $s \neq O(t)$, and so $s
\neq O(t)$, as required. \proofend

%=========================================================================

\section{Handling an increasing sequence of functions}
\label{sequence:section}

We now strengthen the result from the previous section. Write $f \prec
g$ if $f = O(g)$ and $g \neq O(f)$. Let $F$ be the set of functions
from a set $S$ to the nonnegative part of an ordered ring $R$, and
suppose $G_1, \ldots, G_k, G_*$ are any nonnegative functions
satisfying
\[
0 \prec G_1 \prec G_2 \prec \ldots \prec G_k \prec G_*
\]
Suppose we expand our language with function symbols $g_1, \ldots,
g_k$, intended to denote $G_1, \ldots, G_k$. We will now show that
when we are dealing with Horn clauses and the function variables are
assumed to range over $F$, once again, the obvious strategy for
testing validity turns out, surprisingly, to be complete. In this
case, the functions that take negative values and arbitrary
quantifier-free formulas requires some additional hypotheses. We will
therefore deal with the simpler case first.

\begin{theorem}
\label{last:theorem}
Fix $S$, $R$, $F$, and $G_1, \ldots, G_k$ as above. A Horn clause
$\ph$ is valid when the variables range over $F$ and $g_1, \ldots,
g_k$ are interpreted as $G_1, \ldots, G_k$, respectively, if and only
if
\begin{equation}
\label{many2:eq}
0 \prec g_1 \prec g_2 \prec \ldots \prec g_k \limplies \ph
\end{equation}
if valid in the sense of Theorem~\ref{main:theorem}.
\end{theorem}
Thus, we can decide the validity of big O entailments relative to any
sequence of nonnegative functions with strictly increasing rate of
growth, and the results do not depend on which ones we use. Now,
suppose $g_\alpha$ is any set of symbols indexed by a computable
linear ordering $I$. Since any formula can only use finitely many of
them, we have the following:
\begin{corollary}
  Let $F$ be any set of functions from an infinite set $S$ to an the
  nonnegative part of an ordered ring, $R$. Let $\{G_\alpha\}$ be any
  set of functions in $F$, indexed by a computable linear ordering
  $I$, such that $G_\alpha \prec G_\beta$ whenever $\alpha <
  \beta$. Consider the language $L'$ with constants $g_\alpha$ to
  denote the functions $G_\alpha$. Then the set of Horn clauses valid
  in the structure $\la F, \ldots, G_\alpha, \ldots \ra$ is decidable,
  and does not depend on the structure chosen.
\end{corollary}

Clearly if formula~(\ref{many2:eq}) of Theorem~\ref{last:theorem} is
valid, then $\ph$ is valid when $g_1, \ldots, g_k$ are interpreted as
$G_1, \ldots, G_k$. We need to show the converse, i.e.~that of
formula~(\ref{many2:eq}) is false, we can construct a counterexample
to $\ph$ with the same interpretations of $g_1, \ldots, g_k$. The
following lemma will facilitate our task.
\begin{lemma}
\label{last:lemma}
Let $\ph$ by any quantifier-free formula in $L$. Let $f$ and $g$ be
any variables occurring in $\ph$. Then $\ph$ is valid if and only if
the formula
\[
(f = O(g) \lor g = O(f)) \limplies \ph
\]
is valid.
\end{lemma}
Note that here we are dealing with formulas in $L$, not $L'$, and
validity in the sense of Theorem~\ref{main:theorem}. The proof is
virtually identical to that of Lemma~\ref{case:split}: given any
interpretations for $f$ and $g$, we can divide the domain $S$ into the
set $S_0$ on which $|f(x)| \leq |g(x)|$, and the complementary set
$S_1 = S - S_0$.

\medskip

\proofarg{of Theorem~\ref{last:theorem}} 
Let $\ph$ be a Horn clause in the language $L'$, of the form 
\[
\bigwedge q_i = O(r_i) \limplies s = O(t).
\]
Formula~(\ref{many2:eq}) is equivalent to
\begin{multline}
\label{many2b:eq}
\bigwedge g_i = O(g_{i+1}) \land \bigwedge q_j = O(r_j) \limplies\\
g_1 = O(0) \lor g_2 = O(1) \lor \ldots \lor g_k = O(g_{k-1}) \lor s =
O(t).
\end{multline}
On the assumption that this is not valid, we need to construct a
counterexample with the desired interpretations of $g_1, \ldots,
g_k$. We can introduce new variables to name $s$ and $t$, and so
assume without loss of generality that $s$ and $t$ are variables
themselves. Using Lemma~\ref{last:lemma}, we can assume that for
every pair of variables $f$ and $g$, either $f = O(g)$ or $g =
O(f)$ are among the hypotheses of $\ph$.

With this useful simplification, the argument now follows a line of
reasoning similar to that used in
Section~\ref{constants:section}. Since formula~(\ref{many2b:eq}) is
not valid, running the algorithm on each of the $k+1$ disjuncts
returns ``false.'' From the first $k$ runs of the algorithm we get
sets of variables
\[
A_0 \subseteq A_1 \subseteq \ldots \subseteq A_{k-1}, 
\] 
where a variable $f$ is in $A_0$ if and only if $f = 0$ is a consequence
of the hypotheses, and for $i=1,\ldots,k-1$ a variable $f$ is in $A_i$
if and only if $f = O(g_i)$ is a consequence of the hypotheses. In
particular, for $i=1,\ldots,k-1$, $g_i$ is in $A_i$ but not $A_{i-1}$.
We also get assignments $c^0, \ldots, c^{k-1}$ of rational numbers to
the variables in such a way that for each $i$:
\begin{itemize}
\item the assignment $c^i$ satisfies all the hypotheses;
\item $c^i$ assigns $0$ to variables in $A_i$; and
\item $c^i$ assigns strictly positive values to variables not in $A_i$.
\end{itemize}
For notational uniformity, we tack one more set onto the end of the
sequence: let $A_k$ be the set of all the variables in $\ph$, and let
$c^k$ be the assignment that assigns $0$ to every variable.

From the last run of the algorithm we get a set of variables $B$ that
includes $t$ but not $s$, and an assignment $d$ to the variables such
that:
\begin{itemize}
\item $d$ satisfies all the hypotheses;
\item $d$ assigns a value of $0$ to all the variables
  in $B$; and
\item $d$ assigns a strictly positive values to variables not
  in $B$.
\end{itemize}

Now there are three possibilities. Either $B$ contains $0$ but not
$g_1$, or for some $i=1,\ldots,k-1$, $B$ contains $g_i$ but not
$g_{i+1}$, or $B$ contains $g_i$ for every $i$. By the assumption that
$\ph$ fixes an ordering on the rates of growth of the variables, in
the first case, we have $B \subseteq A_1$; in the second case, we have
$A_{i-1} \subseteq B \subseteq A_{i+1}$; in the last case, we have
$A_{k-1} \subseteq B$. In the first case, replace $A_0$ by $B$ and the
assignment $c^0$ by $d$; in the second case, replace $A_i$ by $B$ and
the assignment $c^i$ by $d$; in the third case, replace $A_k$ by $B$
and the assignment $c^k$ by $d$. Then the sets
\[
A_0 \subseteq A_1 \subseteq \ldots \subseteq A_k
\]
and the assignments $c^0, c^1, \ldots, c^k$ have the following
properties:
\begin{itemize}
\item $0$ is in $A_0$, but not $A_1$.
\item For each $i=1,\ldots, k$, $g_i$ is in $A_i$, but not $A_{i-1}$.
\item For some $i < k$, $t$ is in $A_i$, and $s$ is not in $A_i$.
\item For each $i = 0,\ldots,k$:
\begin{itemize}
\item $c^i$ assigns a value of $0$ to all variables in $A_i$;
\item $c^i$ assigns a strictly positive values to variables not in
  $A_i$;
\item $c^i$ satisfies all the hypotheses $q = O(r)$ of $\ph$; in other
  words, $q = q[A_i] = 0$ whenever $r$ is $0$ under the assignment.
\end{itemize}
\end{itemize}
We will assign functions to the variables $f_1, \ldots, f_k, g_1,
\ldots, g_m$ so that:
\begin{itemize}
\item for each $i=1,\ldots,m$, $g_i$ is assigned the value $G_i$;
\item each variable in $A_0$ is assigned $0$;
\item for each $i=1,\ldots,k$, each variable $f$ in $A_i$ but not
  $A_{i-1}$ is a assigned a function that is $O(G_i)$ but not
  $O(G_{i-1})$;
\item each variable not in $A_k$ is assigned a function
  that is $O(G_*)$ but not $O(G_k)$; and
\item all the hypotheses of $\ph$ are satisfied.
\end{itemize}
These conditions imply that for some $i$, $t = O(G_i)$ but $s \neq
O(G_i)$, so $s \neq O(t)$ under the assignment, as required.

Let $H_1, \ldots, H_k$ be functions from $S$ to $R$ having the same
rate of growth as $G_1, \ldots, G_k$. For the moment, this is all we
assume about $H_1, \ldots, H_k$; we will choose particular values for
these functions soon. For each assignment $c^i$, let $c^i(f)$ denote
the rational number assigned to the variable $f$. To each variable $f$,
we assign the function
\[
c^0(f) H_1 + c^1(f) H_2 + c^2(f) H_3 + \ldots + c^{k-1} (f) H_k +
c^k(f) G_*.
\]
It has not hard to see that this assignment gives the variables the
orders of growth claimed.

Let us show that the hypotheses of $\ph$ are satisfied under the
assignment. Let $q = O(r)$ be one of these hypotheses. If $r$ has a
function symbol that is not in $A_k$, then $G_* = O(r)$, and $q =
O(r)$ is satisfied immediately. Otherwise, let $i$ be the largest
index such that $r$ has a variable in $A_i$. Then $H_i = O(r)$, and
all that matters are the coefficients of $H_{i+1}, \ldots, H_k, G_*$
in $q$; in other words, all that matters are the coefficients of
$q[A_i]$. But since all of the variables of $r$ are in $A_i$, the
assignments $c^{i+1}, \ldots, c^k$ were chosen to ensure that all the
coefficients of $H_{i+1}, \ldots, H_k, G_*$ in $q[A_i]$ are $0$, as
required.

We only need to choose $H_1, \ldots, H_k$ so that $g_1, \ldots,
g_k$ receive the values $G_1, \ldots, G_k$. But because, for each $i$,
$g_i$ is in $A_i$ but not $A_{i-1}$, $g_i$ is assigned a value of the form
\[
a_{i,1} H_1 + a_{i,2} H_2 + \ldots + a_{i,i} H_i,
\]
where each coefficient is strictly positive. Set each of these values
to the corresponding $G_i$; now it is not hard to see that we can
iteratively solve for $H_i$ in terms of $G_i$, and that each $H_i$
will be an expression involving $G_1, \ldots, G_i$ in which $G_i$ has
a nonzero coefficient. Thus, for this choice of $H_1, \ldots, H_k$,
all the conditions are satisfied, and we have the desired
counterexample. \proofend

In order to extend the decision procedure above to arbitrary
quantifier-free formulas, we need to be able to combine
counterexamples, as in the discussion at the beginning of
Section~\ref{procedure:section}. And to extend the decision procedure
to functions that also take negative values, we need an analogue to
Lemma~\ref{case:split}, whose proof relied on the ability to extend a
counterexample on a subset of the domain. Once we fix functions $G_1,
G_2, \ldots, G_k$, however, both these requirements are problematic,
unless we impose further assumptions. For example, the assertion $G_i
\prec G_j$ only describes the global behavior of $G_i$ and $G_j$,
leaving the possibility that additional information is encoded in the
set of values $x$ where $G_j(x) < G_i(x)$.

We will henceforth assume that $S$ carries a linear ordering, $<$, and
has no greatest element. If $A$ and $B$ are subsets of $S$, we will
say that $A$ is \emph{cofinal in $B$} if for every $b$ in $B$, there
is an $a$ in $A$ such that $a \geq b$. Note that if $A$ is cofinal in
$B$ and $B$ is cofinal in $C$ then $A$ is cofinal in $C$, and any set
cofinal in $S$ is infinite. We now impose the following additional
restrictions:
\begin{itemize}
\item We read $f = O(g)$ as the assertion that $f$ is
  \emph{eventually} $O(g)$, that is, for some $C$ and $y \in S$ we
  have $\fa {x > y} (|f(x)| \leq C |g(x)|)$.
\item We assume that the relationships
\[
0 \prec G_1 \prec G_2 \prec \ldots \prec G_k \prec G_*
\]
also hold of the restrictions of the $G_i$'s to any cofinal subset of
$S$.
\end{itemize}
The second clause says that, in a sense, the relationships between the
$G_i$'s is robust. This clause is clearly satisfied by the functions
given in the example in Section~\ref{introduction:section}. We now
show that these restrictions are enough to ensure that an analogue of
Lemma~\ref{case:split} holds for Horn clauses.

\begin{lemma}
  Let $S$ and $G_1, \ldots, G_k, G_*$ be as above, and let $\ph(f)$ be
  any Horn clause in $L'$. Then $\ph(f)$ is valid for interpretations
  where the variables range over functions from any cofinal subset of
  $S$ to $R$, and $g_1, \ldots, g_k$ are interpreted as the
  corresponding restrictions of $G_1, \ldots, G_k$, respectively, if
  and only if $\ph(\alpha)$ and $\ph(-\alpha)$ are both valid for the
  same class of interpretations, with $\alpha$ restricted to range
  over nonnegative functions.
\end{lemma}

\proof
  If $\ph(f)$ is valid, then so are $\ph(\alpha)$ and $\ph(-\alpha)$,
  so we only need to prove the converse. Let $\ph(f)$ be the Horn
  clause $\bigwedge q_i = O(r_i) \limplies s = O(t)$, and suppose $\ph(f)$
  is not valid. Fix a counterexample, which therefore makes each
  equation $q_i = O(r_i)$ true and $s = O(t)$ false for some cofinal
  subset $S'$ of $S$. Let $S_0$ be the set of elements $x$ in $S'$ such
  that $f(x)$ is nonnegative, and let $S_1 = S' - S_0$. The fact that
  $s = O(t)$ is false on $S'$ means that it is false of the
  restriction to either $S_0$ or $S_1$. Since we are
  using the ``eventually'' reading of big O, we can further assume
  that this $S_i$ is cofinal in $S'$, and hence cofinal in $S$. Thus
  we have a counterexample to the validity of $\ph(\alpha)$ or a
  counterexample to the validity of $\ph(-\alpha)$, as required.
\proofend

By the reductions in Section~\ref{negative:section}, we therefore have
the following:

\begin{theorem}
  Given $S, R, F$, and $G_1, \ldots, G_k$ as above and the
  ``eventually'' reading of the big O relation, the set of Horn
  clauses of $L'$ valid in this interpretation is decidable.
\end{theorem}

Recall that every ordered ring $R$ contains a copy of the natural
numbers. To extend our decision procedure to arbitrary formulas, we
impose the following additional restrictions:
\begin{itemize}
\item The image of $\N$ is cofinal in $R$.
\item There is a countable cofinal subset of $S$. 
\end{itemize}
Note that both these restrictions hold when $R$ and $S$ are any of the
sets $\Z$, $\Q$, or $\R$.

\begin{lemma}
  With the additional assumptions above, any formula $\ph$ of $L'$ of
  the form
\[
\bigwedge q_i = O(r_i) \limplies \bigvee_{j=1}^m s_j = O(t_j)
\]
is valid if and only if each formula 
\[
\bigwedge q_i = O(r_i) \limplies s_j = O(t_j)
\]
is valid for each $j=1,\ldots,m$.
\end{lemma}

\proof Suppose we are given a counterexample to the formula $\bigwedge
q_i = O(r_i) \limplies s_j = O(t_j)$ for each $j = 1,\ldots,m$. We
need only show how to amalgamate these counterexamples. Since $s_j =
O(t_j)$ is false in the $j$th counterexample for each $j=1,\ldots,m$,
we can choose, for each $n \in N$, an element $x_{j,n}$ in $S$ such
that
\[
| s_j(x_{j,n}) | > n \cdot |t_j(x_{j,n}) |
\]
is satisfied under that interpretation. Since we are using the
``eventually'' reading of big O and assuming there is a countable
cofinal subset of $S$, we can further assume that for each $j$, the
sequence $\la x_{j,n} \ra_{n \in N}$ is increasing and cofinal in $S$. We
can then thin out these sequences, deleting elements that are
duplicated, to ensure that they are disjoint.

Now define a new interpretation by interpreting each function symbol
according to the $j$th counterexample on the sequence $\la x_{j,n}
\ra_{n \in \N}$, and, say, according to the first counterexample on
all the other elements of $S$. Then this interpretation will still
satisfy $q_i = O(r_i)$ for each $i$, since each of the counterexamples
does. But for each $j$, we have guaranteed that $s_j = O(t_j)$ is
false, since for any $y$ in $S$ and $n$ in $\N$ we have guaranteed
that $|s_j(x)| > n \cdot |t_j(x)|$ for some $x > y$.  \proofend

Thus, by the observations at the beginning of
Section~\ref{procedure:section}, we can extend decidability from Horn
clauses to arbitrary quantifier-free formulas. 

\begin{theorem}
  Given $S, R, F$, and $G_1, \ldots, G_k$ satisfying the additional
  restrictions above, and the ``eventually'' reading of the big O
  relation, the set of quantifier-free formulas of $L'$ valid in this
  interpretation is decidable.
\end{theorem}

%=========================================================================

\section{Questions}
\label{questions:section}

There are a number of interesting theoretical puzzles, as well
interesting pragmatic challenges, that remain.

We have restricted our attention to linear terms. A number of useful
big O identities hold of terms involving multiplication and
composition of functions (see
\cite{avigad:donnelly:04,graham:et:al:94}). We do not know, for
example, whether the quantifier-free fragment of the language is
decidable in the presence of multiplication. Nor do we know whether
anything useful can be said about composition. 

Our handling of constant functions in Section~\ref{constants:section}
presupposed that the range of the set of functions is an ordered
field. We do not know, for example, whether the linear theory of big O
equations involving functions from $\N$ to $\Z$ is decidable when we
include the constant function $1$, or even whether the set of
validities described in Section~\ref{constants:section} is complete.

We also do not know whether the full first-order theory of the linear
fragment of big O reasoning is decidable. In practice, however, this
theory does not seem to be very useful.

Even in cases where the full theory is undecidable, we suspect that
there are reasonable procedures that capture most of the inferences
that come up in practice, and do so efficiently. We are fortunate that
the simple decision procedure we provide here seems to be
pragmatically useful as well. In general, although clean decidability
and undecidability results provide a useful sense of what can be done
in principle, when it comes to formal verification, it is equally
important to find principled approaches to developing imperfect
methods that work well in practice. (See, for example,
\cite{avigad:friedman:unp} for a study of heuristic procedures for
inequalities between real valued expressions that is motivated by this
philosophy.)

% \newcommand{\nameindex}[1]{}

% \bibliographystyle{plain} 
% \bibliography{proofthry,automated}

\begin{thebibliography}{10}

\bibitem{apt:03}
Krzysztof Apt.
\newblock {\em Principles of constraint programming}.
\newblock Cambridge University Press, Cambridge, 2003.

\bibitem{avigad:donnelly:04}
Jeremy Avigad and Kevin Donnelly.
\newblock Formalizing {O} notation in {I}sabelle/{HOL}.
\newblock In David Basin and Micha\"el Rusinowitch, editors, {\em Automated
  reasoning: second international joint conference, IJCAR 2004}, 
  Springer-Verlag, 2004, 357--371.

\bibitem{avigad:et:al:unp}
Jeremy Avigad, Kevin Donnelly, David Gray, and Paul Raff.
\newblock A formally verified proof of the prime number theorem.
\newblock To appear in \emph{ACM transactions on computational logic}.

\bibitem{avigad:friedman:unp}
Jeremy Avigad and Harvey Friedman.
\newblock Combining decision procedures for the reals.
\newblock \emph{Logical methods in computer science}, 2(4:4), 2006

\bibitem{bertot:casteran:04}
Yves Bertot and Pierre Cast\'eran.
\newblock Interactive theorem proving and program development:
Coq'art: the calculus of inductive constructions. 
\newblock Springer-Verlag, 2004.

\bibitem{gordon:melham:93}
Michael J.~C.~Gordon and Thomas F.~Melham.
\newblock {\em Introduction to HOL: a theorem proving environment for
  higher-order logic}.
\newblock Cambridge University Press, 1993. 

\bibitem{graham:et:al:94}
Ronald~L. Graham, Donald~E. Knuth, and Oren Patashnik.
\newblock {\em Concrete mathematics: a foundation for computer science}.
\newblock Addison-Wesley Publishing Company, Reading, MA, second
edition, 1994.

\bibitem{hall:86}
Marshall Hall, Jr.
\newblock {\em Combinatorial theory}.
\newblock John Wiley \& Sons Inc., New York, second edition, 1986.

\bibitem{karmarkar:84}
Narendra Karmarkar.
\newblock A new polynomial time algorithm for linear programming.
\newblock {\em Combinatorica}, 4:373--395, 1984.

\bibitem{kaufmann:et:al:00}
Matt Kaufmann, Panagiotis Manolios, and J.~Strother Moore.
\newblock \emph{Computer-aided reasoning: an approach}.
\newblock Kluwer Academic Publishers, 2000.

\bibitem{loos:weispfenning:93} 
R\"udiger Loos and Volker Weispfenning. 
\newblock Applying linear quantifier elimination. 
\newblock {\em The computer journal}, 36:450--461, 1993.

\bibitem{nipkow:et:al:02}
Tobias Nipkow, Lawrence~C. Paulson, and Markus Wenzel.
\newblock {\em Isabelle/{HOL}: a proof assistant for higher-order logic}.
\newblock Springer-Verlag, Berlin, 2002.

\bibitem{owre:et:al:92}
S.~Owre, J.~M.~Rushby, and N.~Shankar.
\newblock PVS: a prototype verification system.
\newblock In \emph{Proceedings of the 11th international conference on
  automated deduction (CADE)}, Springer-Verlag, 1992, 748--752.

\bibitem{papadimitriou:steiglitz:98}
Christos~H. Papadimitriou and Kenneth Steiglitz.
\newblock {\em Combinatorial optimization: algorithms and complexity}.
\newblock Dover Publications Inc., Mineola, NY, 1998.
\newblock Corrected reprint of the 1982 original, Prentice-Hall, New Jersey.

\bibitem{weispfenning:88}
Volker Weispfenning. 
\newblock The complexity of linear problems in fields. 
\newblock {\em Journal of symbolic computation}, 5:3--27, 1988.

\bibitem{weispfenning:94}
Volker Weispfenning. 
\newblock Parametric linear and quadratic optimization by elimination.
\newblock Technical report MIP-9404, Universit\"at Passau, 1994. 

% \bibitem{coq}
% The {C}oq proof assistant.
% \newblock Developed by the LogiCal project.\\
%   http://pauillac.inria.fr/coq/coq-eng.html.

% \bibitem{pvs}
% The PVS specification and verification system.
% \newblock http://pvs.csl.sri.com/.

\end{thebibliography}

\end{document}